\documentclass{desyproc}

\begin{document}
\title{Phenomenology of Majorons}

\author{{\slshape Julian Heeck}\\[1ex]
Service de Physique Th\'eorique, Universit\'e Libre de Bruxelles, CP225, 1050 Brussels, Belgium
}

\contribID{4}

\confID{13889}  
\desyproc{DESY-PROC-2017-XX}
\acronym{Patras 2017} 
\doi  

\maketitle

\begin{abstract}
Majorons are the Goldstone bosons associated to lepton number and thus closely connected to Majorana neutrino masses. Couplings to charged fermions arise at one-loop level, including lepton-flavor-violating ones that lead to decays $\ell\to \ell' J$, whereas a coupling to photons is generated at two loops. The typically small couplings make massive majorons a prime candidate for long-lived dark matter. Its signature decay into two mono-energetic neutrinos is potentially detectable for majoron masses above MeV.
\end{abstract}

\section{Majoron couplings}

The difference between baryon number $B$ and lepton number $L$ is an anomaly-free global symmetry of the Standard Model (SM); spontaneously breaking this $U(1)_{B-L}$ symmetry results in a Goldstone boson called majoron~\cite{Chikashige:1980ui,Schechter:1981cv}. In the simplest realization, this majoron $J$ resides in a singlet complex scalar $\sigma = (f+ \sigma^0 + i J)/\sqrt{2}$ that carries $B-L$ charge 2, $f$ being the $B-L$ breaking scale and $\sigma^0$ the heavy CP-even majoron partner. Further introducing three right-handed neutrinos~$N_R$, the Lagrangian reads
\begin{align}
\mathcal{L} = \mathcal{L}_\text{SM}+ i\overline{N}_R \gamma^\mu\partial_\mu N_R + (\partial_\mu \sigma)^\dagger (\partial^\mu \sigma) - V(\sigma)-\left(\overline{L} y N_R H +\tfrac12\overline{N}_R^c\lambda  N_R \sigma +\text{h.c.}\right) ,
\label{eq:lagrangian}
\end{align}
with the SM lepton (scalar) doublet $L$ ($H$). We suppressed flavor indices and the details of the scalar potential $V(\sigma)$. $SU(2)_L\times U(1)_Y\times U(1)_{B-L}$ symmetry breaking then yields the famous seesaw neutrino mass matrix $M_\nu \simeq - m_D M_R^{-1} m_D^T$ with $m_D = y v/\sqrt{2}$ and $M_R = \lambda f/\sqrt{2} \gg m_D$.

Many of the parameters encoded in $M_\nu$ have been measured already: the mass splittings and mixing angles. However, even if we could measure all elements of $M_\nu$, we would still not be able to reconstruct the underlying seesaw parameters $m_D$ and $M_R$. As shown in Ref.~\cite{Davidson:2001zk}, one can map the parameters $\{m_D, M_R\}$ bijectively onto $\{M_\nu, m_D m_D^\dagger\}$, implying that $m_D m_D^\dagger$ contains precisely those nine seesaw parameters that cannot be determined by measurements of neutrino masses and oscillations. As we will see below, this is a convenient parametrization for the phenomenology of majorons, which endow $m_D m_D^\dagger$ with physical meaning.

The tree-level couplings of the majoron $J$ can easily be derived from Eq.~\eqref{eq:lagrangian}, which in particular include the couplings $J \overline{\nu}_j i \gamma_5 \nu_j m_j/(2f)$ to the light neutrino mass eigenstates $\nu_j$. With $f$ at the seesaw scale and active neutrino masses $m_j$ below eV, this coupling is incredibly tiny. At one-loop level~\cite{Chikashige:1980ui,Pilaftsis:1993af,Garcia-Cely:2017oco}, the majoron also obtains couplings to charged leptons $\ell$ and quarks~$q$, parametrized as $i J \bar{f}_1 (g_{Jf_1 f_2}^S+ g_{Jf_1 f_2}^P \gamma_5) f_2$ with coefficients
\begin{align}
g_{J q q'}^P &\simeq \frac{m_q }{8\pi^2 v}\delta_{q q'} T^q_3 \, \text{tr} K\,, & 
g_{J q q'}^S &= 0 \,,\\
g_{J\ell{\ell'}}^P &\simeq \frac{m_\ell+m_{\ell'}}{16\pi^2 v} \left(\delta_{\ell \ell'} T^\ell_3\, \text{tr} K +  K_{\ell\ell'} \right), & 
g_{J\ell{\ell'}}^S &\simeq \frac{m_{\ell'}-m_\ell}{16\pi^2 v}  K_{\ell\ell'} \,,
\label{eq:fermion_couplings}
\end{align}
where $T^{d,\ell}_3= - T^u_3 = -1/2$ and  we introduced  the dimensionless hermitian coupling matrix $K \equiv  m_D m_D^\dagger/(v f)$. The majoron couplings to charged fermions are hence determined by the seesaw parameters $m_D m_D^\dagger$, which are \emph{independent} of the neutrino masses and can in particular be much bigger than the naive one-generation expectation $M_\nu M_R$. Perturbativity sets an upper bound on $K$ of order $4\pi v/f$, and since $K$ is furthermore positive definite we have the inequalities $|K_{\ell\ell'}|\leq \sqrt{K_{\ell\ell} K_{\ell'\ell'}} \leq \text{tr} K$. These fermion couplings are obviously crucial for majoron phenomenology and in principle even offer a new avenue to reconstruct the seesaw parameters. Note in particular the off-diagonal lepton couplings, which will lead to lepton flavor violation~\cite{Pilaftsis:1993af,Garcia-Cely:2017oco} (Sec.~\ref{sec:lfv}).

There is one more coupling of interest, that to photons. For a massless majoron, the coupling $J F \tilde{F}$ vanishes because $B-L$ is anomaly free~\cite{Pilaftsis:1993af}; otherwise, it is induced at two-loop level and non-trivial to calculate. Considering only a gauge-invariant subset of diagrams, we can however obtain the simple expression~\cite{Garcia-Cely:2017oco}
\begin{align}
\Gamma (J\to\gamma\gamma) \simeq \frac{\alpha^2 \left(\text{tr}K\right)^2}{4096\pi^7} \frac{m_J^3}{v^2}  \left| \sum_f  N_c^f T_3^f Q_f^2 \,g \left(\frac{m_J^2}{4 m_f^2}\right)\right|^2 ,
\label{eq:photon_couplings}
\end{align} 
where the sum is over all SM fermions $f$ with color multiplicity $N_c^f$, isospin $T_3^f$, and electric charge $Q_f$. The loop function is given by $g(x) \equiv - (\log [1-2 x + 2\sqrt{x (x-1)}])^2/(4 x)$.

\section{Majoron dark matter}

With the relevant majoron couplings at our disposal, we can start to discuss phenomenology. First off, we are going to study the case of the majoron as a dark matter (DM) candidate. This is motivated by the fact that it generically has tiny couplings to the SM, ensuring that it is dark and stable enough to form DM~\cite{Rothstein:1992rh,Berezinsky:1993fm}. A prerequisite here is an explicit $U(1)_{B-L}$ breaking in the Lagrangian to generate a majoron mass $m_J$, making $J$ a \emph{pseudo}-Goldstone boson. This could simply be an explicit mass term in the scalar potential, a gravity-generated higher-dimensional operator or an axion-like anomaly-induced potential. Furthermore, a production mechanism is required to generate the observed abundance in the early Universe. With small couplings, the obvious mechanism to use here is \emph{freeze-in}, e.g.~from the coupling to the Higgs or the right-handed neutrinos~\cite{Frigerio:2011in}. For majoron masses as low as keV one has to be careful not to violate structure-formation constraints from the Lyman-$\alpha$ forest. In these cases, different production mechanisms are required that make $J$ cold enough, which can naturally be found in inverse-seesaw majoron models~\cite{Heeck:2017xbu,Boulebnane:2017fxw}. Here we will focus on DM masses above MeV for simplicity.

Assuming a massive singlet majoron to make up all of DM, the main signature then comes from its eventual decay into SM particles. As discussed above, the only decay channel at tree level is into neutrino mass eigenstates, $J\to \nu_j\nu_j$, with coupling $m_j/f$. 
These neutrinos will not oscillate, so the flavor content of the monochromatic neutrino flux follows simply from the mass eigenstates~\cite{Garcia-Cely:2017oco}. 
For normal hierarchy, this implies only a small $\nu_e$ component of the flux, because the heaviest neutrino only has a tiny $\theta_{13}$-suppressed electron component; for inverted hierarchy, the majoron decays into the two heaviest neutrinos, which results in roughly 50\% electron flavor in the flux; in the quasi-degenerate regime, all flavors are equally probable.

\begin{wrapfigure}{r}{0.62\textwidth}
\centerline{\includegraphics[width=0.6\textwidth]{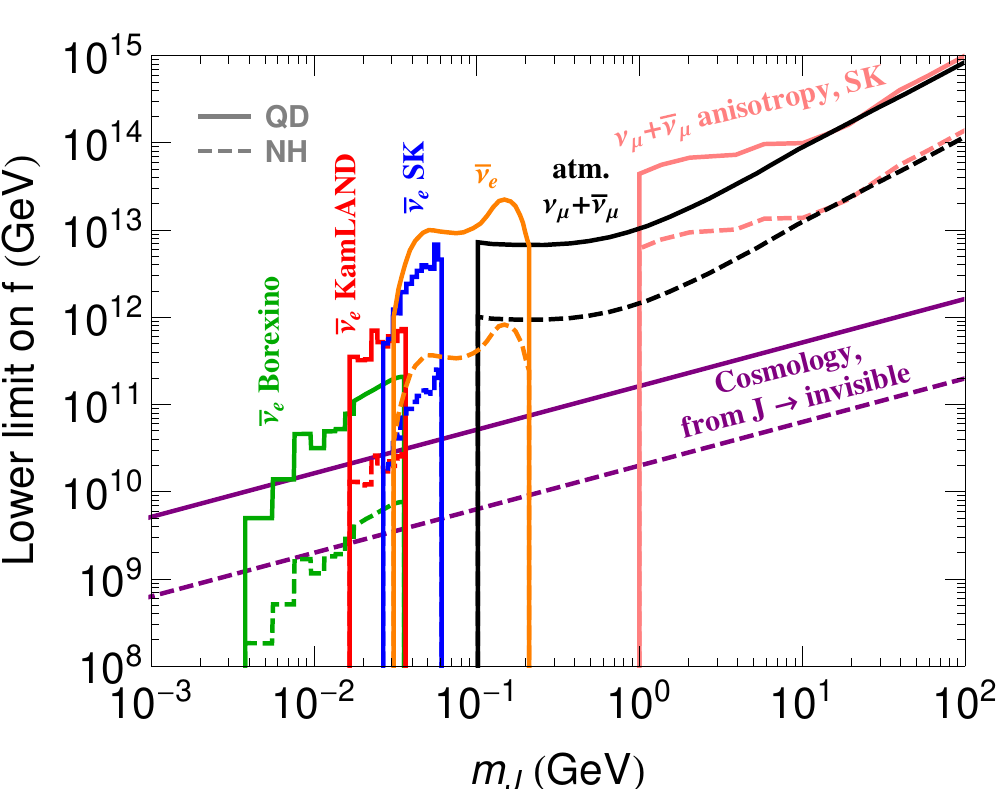}}
\caption{Lower limit on the $B-L$ breaking scale $f$ from DM decay $J\to \nu\nu$, assuming a quasi-degenerate (solid) or normal-hierarchy neutrino spectrum (dashed)~\cite{Garcia-Cely:2017oco}.}
\label{Fig:Heeck_Julian_fig1}
\end{wrapfigure}
Knowing the flavor composition of $J\to \nu\nu$ allows us to search for these neutrino lines with neutrino detectors. Borexino and KamLAND use inverse beta decay $\overline{\nu}_e p \to n e^+$ to reconstruct the neutrino energy with good accuracy. Due to the kinematic threshold of this process it is not possible to detect neutrino lines below $m_J\sim \text{MeV}$. Above MeV, on the other hand, these experiments could indeed be sensitive to a dark-matter induced neutrino flux~\cite{Garcia-Cely:2017oco} (see Fig.~\ref{Fig:Heeck_Julian_fig1}). For higher masses, Super-K becomes most sensitive and can also utilize the $\nu_\mu$ component of the flux~\cite{PalomaresRuiz:2007ry}. For sub-MeV masses, limits on $J\to\nu\nu$ can still be derived from cosmology~\cite{Poulin:2016nat}, but are of course less of a smoking-gun signature for majoron DM.

Majoron DM can thus be used to motivate neutrino line searches all the way down to MeV energies, far below what is typically considered. A natural question to ask here is whether observable neutrino fluxes are compatible with limits from visible DM decay channels, which are far more constrained. As shown above, the decays $J\to \ell \bar{\ell}', q\bar{q}, \gamma\gamma$ are indeed all unavoidably induced at loop level in the singlet majoron model. However, they all depend on parameters that are independent of the $J\to\nu\nu$ channel, making it impossible to directly compare these channels. In other words, the DM decay into visible channels probes different parameters than $J\to \nu\nu$, making them \emph{complementary}. In the $m_J=\text{MeV}$--$100\,\text{GeV}$ region, one can indeed obtain strong constraints on the $K$ matrix elements from the visible channels, without invalidating our conclusion about neutrino lines~\cite{Garcia-Cely:2017oco}.
For sub-MeV majoron masses, only the decay $J\to\gamma\gamma$ remains as a promising indirect detection signature~\cite{Berezinsky:1993fm,Bazzocchi:2008fh}.

\section{Lepton flavor violation}
\label{sec:lfv}

Going back to the majoron couplings to fermions of Eq.~\eqref{eq:fermion_couplings} shows that the quark couplings are diagonal at one-loop level, whereas the lepton couplings are not. Due to the rather strong lepton mass hierarchy, $m_\ell \gg m_{\ell'}$, the off-diagonal couplings can be approximately written as $ -\frac{i m_\ell}{8\pi^2 v} K_{\ell {\ell'}}\, J \,\bar{\ell} P_L {\ell'} + \text{h.c.} $, which can induce the lepton-flavor-violating two-body decays $\ell\to \ell' J$~\cite{Pilaftsis:1993af,Garcia-Cely:2017oco}. If the majoron is massless or decays invisibly, the only signature of this decay is the mono-energetic $\ell'$, which has to be searched for on top of the continuous energy spectrum from the SM decay channel $\ell\to \ell' \nu_{\ell} \overline{\nu}_{\ell'}$. Current limits translate into $|K_{\mu e}|\lesssim 10^{-5}$, $|K_{\tau \ell}|\lesssim \mathcal{O}(10^{-3})$, with good prospects for improvement at Mu3e and Belle~\cite{Heeck:2016xkh,Yoshinobu:2017jti}. Channels with more tagging potential, such as $\ell\to \ell' J\gamma$ or $\ell \to \ell' (J\to \text{visible})$, are also promising and currently under investigation.
We stress that lepton flavor violation with majorons depends on a different combination of seesaw parameters than the more commonly studied heavy-neutrino induced $\ell\to \ell' \gamma$. These channels are therefore complementary and should both be investigated.

\section{Conclusion}

The singlet majoron model inherits some nice properties from the seesaw Lagrangian, namely small Majorana neutrino masses and leptogenesis, while providing a new phenomenological handle. The majoron couplings to charged particles are precisely given by the seesaw parameters that are impossible to determine from the neutrino mass matrix, which could in principle allow us to reconstruct the seesaw with low-energy measurements. Since the couplings can be tiny without fine-tuning, a massive majoron makes for a promising unstable dark matter candidate, with signature decay into mono-energetic neutrinos, potentially detectable for energies above MeV.
With few new parameters, which are furthermore linked to the seesaw mechanism, majoron models are simple extensions of the Standard Model that still provide rich phenomenology.

\section*{Acknowledgments}
JH is a postdoctoral researcher of the F.R.S.-FNRS and thanks the PATRAS organizers for an interesting conference and Camilo Garcia-Cely for a fruitful collaboration.


\begin{footnotesize}

\end{footnotesize}



\begin{thebibliography}{99}
%


\bibitem{Chikashige:1980ui} 
  Y.~Chikashige, R.~N.~Mohapatra and R.~D.~Peccei,
  Phys.\ Lett.\  {\bf 98B}, 265 (1981).



\bibitem{Schechter:1981cv} 
  J.~Schechter and J.~W.~F.~Valle,
  Phys.\ Rev.\ D {\bf 25}, 774 (1982).



\bibitem{Davidson:2001zk} 
  S.~Davidson and A.~Ibarra,
  JHEP {\bf 0109}, 013 (2001)
  [hep-ph/0104076].



\bibitem{Pilaftsis:1993af} 
  A.~Pilaftsis,
  Phys.\ Rev.\ D {\bf 49}, 2398 (1994)
  [hep-ph/9308258].



\bibitem{Garcia-Cely:2017oco} 
  C.~Garcia-Cely and J.~Heeck,
  JHEP {\bf 1705}, 102 (2017)
  [arXiv:1701.07209 [hep-ph]].



\bibitem{Rothstein:1992rh} 
  I.~Z.~Rothstein, K.~S.~Babu and D.~Seckel,
  Nucl.\ Phys.\ B {\bf 403}, 725 (1993)
  [hep-ph/9301213].



\bibitem{Berezinsky:1993fm} 
  V.~Berezinsky and J.~W.~F.~Valle,
  Phys.\ Lett.\ B {\bf 318}, 360 (1993)
  [hep-ph/9309214].



\bibitem{Frigerio:2011in} 
  M.~Frigerio, T.~Hambye and E.~Masso,
  Phys.\ Rev.\ X {\bf 1}, 021026 (2011)
  [arXiv:1107.4564 [hep-ph]].



\bibitem{Heeck:2017xbu} 
  J.~Heeck and D.~Teresi,
  Phys.\ Rev.\ D {\bf 96}, 035018 (2017)
  [arXiv:1706.09909 [hep-ph]].

\bibitem{Boulebnane:2017fxw}
  S.~Boulebnane, J.~Heeck, A.~Nguyen and D.~Teresi,
  arXiv:1709.07283 [hep-ph].


\bibitem{PalomaresRuiz:2007ry} 
  S.~Palomares-Ruiz,
  Phys.\ Lett.\ B {\bf 665}, 50 (2008)
  [arXiv:0712.1937 [astro-ph]].



\bibitem{Poulin:2016nat} 
  V.~Poulin, P.~D.~Serpico and J.~Lesgourgues,
  JCAP {\bf 1608}, 036 (2016)
  [arXiv:1606.02073 [astro-ph.CO]].



\bibitem{Bazzocchi:2008fh} 
  F.~Bazzocchi, M.~Lattanzi, S.~Riemer-S{\o}rensen and J.~W.~F.~Valle,
  JCAP {\bf 0808}, 013 (2008)
  [arXiv:0805.2372 [astro-ph]].



\bibitem{Heeck:2016xkh} 
  J.~Heeck,
  Phys.\ Lett.\ B {\bf 758}, 101 (2016)
  [arXiv:1602.03810 [hep-ph]].



\bibitem{Yoshinobu:2017jti} 
  T.~Yoshinobu {\it et al.} [Belle Collaboration],
  Nucl.\ Part.\ Phys.\ Proc.\  {\bf 287-288}, 218 (2017).


\end{thebibliography}
\end{document}